\documentclass[12pt]{article}
\usepackage{amsfonts,amssymb,amsmath,graphicx}
\usepackage{epsfig}
\usepackage{epic,eepic}
\usepackage{theorem,cite}
\usepackage{color}

\numberwithin{equation}{section}

\newcommand{\nonu}{\nonumber \\}

\newcommand{\hs}[1]{\hspace{#1 mm}}

                    \def\cF{{\cal F}}
\def\cG{{\cal G}}

                    \def\cR{{\cal R}}
          \def\cT{{\cal T}}          
                    
                    \def\e{{\rm e}} 
\def\ri{{\rm i}}           \def\rd{{\rm d}}           \def\tx{{\widetilde x}}

\def\fA{{\mathfrak A}}

\def\fa{{\mathfrak a}}

\def\fc{{\mathfrak c}}

\def\fr{{\mathfrak r}}
\def\fs{{\mathfrak s}}

\newcommand{\RR}{\mbox{${\mathbb R}$}}

\newcommand{\II}{\mbox{${\mathbb I}$}}

\newcommand{\prt}{\partial}

\newcommand{\wh}[1]{\widehat{#1}}
\newcommand{\wt}[1]{\widetilde{#1}}
\newcommand{\mb}[1]{\hs{4}\mbox{#1}\hs{4}}

\newcommand{\half}{\frac{1}{2}}

\newcommand{\atopn}[2]{\genfrac{}{}{0pt}{}{#1}{#2}}

\definecolor{brique}{rgb}{.9,.2,0}
\definecolor{blvert}{rgb}{0,.8,.85}
\definecolor{vertcl}{rgb}{0,1,.7}
\newcommand\vertcl[1]{\textcolor{vertcl}{#1}}
\newcommand\blvert[1]{\textcolor{blvert}{#1}}
\newcommand\brique[1]{\textcolor{brique}{#1}}
\def\lapth{
\begin{picture}(164,70)(0,-15)\thicklines
\put(0,0){\vertcl{\rule{20pt}{4pt}}}
\put(19,1){\vertcl{\line(1,3){23}}} 
\put(20,1){\vertcl{\line(1,3){23}}} 
\put(21,1){\vertcl{\line(1,3){23}}}
\put(22,1){\vertcl{\line(1,3){23}}}
\put(45,70){\vertcl{\line(1,-3){23}}} 
\put(44,70){\vertcl{\line(1,-3){23}}} 
\put(43,70){\vertcl{\line(1,-3){23}}}
\put(42,70){\vertcl{\line(1,-3){23}}}
\put(2,24){\vertcl{\rule{120pt}{4pt}}}
\put(65,0){\vertcl{\rule{60pt}{4pt}}}
\put(5,37){\Huge{\brique{\textbf{L}}}} 
\put(62,37){\Huge{\brique{\textbf{PTh}}}}
\put(12,-8){\blvert{\rule{92pt}{3.5pt}}}
\put(24,-15){\blvert{\rule{57pt}{3.5pt}}}
\put(36,-22){\blvert{\rule{30pt}{3.5pt}}}
\end{picture}
\raisebox{35pt}{
\begin{minipage}{320pt}\begin{center}
\textbf{Laboratoire d'Annecy-leVieux de Physique
Th\'eorique}\\[4ex]
website: \texttt{http://lappweb.in2p3.fr/lapth-2005/}
\end{center}
\end{minipage}}\\
\vspace{10pt}\quad \hrulefill\\
\vspace{10pt}}

\begin{document}

\pagestyle{empty}
\hspace{-1cm}\lapth

\vfill

\begin{center}
{\Large\bf Algebraic approach to multiple defects on the line\\[1.2ex]
and application to Casimir force}
\\[2.1em]

\bigskip

{\large
M. Mintchev$^{a}$\footnote{mintchev@df.unipi.it} 
and E. Ragoucy$^{b}$\footnote{ragoucy@lapp.in2p3.fr}}\\

\null

\noindent 

{\it 
$^a$ INFN and Dipartimento di Fisica, Universit\`a di
Pisa, Largo Pontecorvo 3, 56127 Pisa, Italy\\[2.1ex] 
$^b$ LAPTH, 9, Chemin de Bellevue, BP 110, F-74941 Annecy-le-Vieux 
cedex, France}
\vfill

\end{center}

\begin{abstract} 

An algebraic framework for quantization in presence of arbitrary number of
point-like defects on the line is developed. We consider a scalar field which
interacts with the defects and freely propagates away of them. As an application 
we compute the Casimir force both at zero and finite temperature. We derive 
also the charge density in the Gibbs state of a complex scalar field with defects.
The example of two delta-defects is treated in detail. 

\end{abstract}
\bigskip 
\medskip 
\bigskip

\vfill
\rightline{IFUP-TH 07/2007}
\rightline{LAPTH-1177/07}
\rightline{\texttt{hep-th/yymm.nnnn}}
\newpage
\pagestyle{plain}
\setcounter{page}{1}


\section{Introduction}

Quantum fields with impurities (defects) are recently 
subject of intensive studies in the context of condensed matter 
physics \cite{KF}-\cite{Fradkin:2006mb}, conformal field theory 
\cite{Saleur:1998hq, Saleur:2000gp} and 
integrable systems \cite{Cherednik:jt}-\cite{Caudrelier:2004hj}. 
Most of the papers concern the case of one 
point-like defect. For more realistic applications 
however, the generalization to $n>1$ defects would be essential. 
In the present paper we face the multi-defect problem by extending 
the algebraic technique developed in 
\cite{Mintchev:2002zd}-\cite{Mintchev:2005rz} for $n=1$. 
We will show below that this technique perfectly applies to  
the case $n>1$ as well, the key point being the reformulation of the boundary
value problem at hand in algebraic terms. 

An immediate application of this formalism is the derivation of the 
Casimir force acting on a test particle. Since this issue is of some 
experimental interest,  we treat it in detail, deriving the general expression of force. 
We consider also a complex scalar field and determine the relative charge density. 
As an explicit example we focus on two delta-type defects. 
It appears that the intensity and direction of the force depend on the position. The 
force is discontinuous at the defects and actually diverges 
when one is approaching any of them from the left (right). A compensation occurs however 
in the symmetric limit, which is finite.  

The paper is organized as follows. 
Introducing the notation, in the next section we briefly summarize 
the algebraic approach to one defect. This approach is generalized 
in section \ref{sec:2defect}, where the solution for two defects is established. 
The latter is extended to $n>2$ defects 
in section \ref{sec:Ndef}. In section \ref{sec:fieldtheo} 
we establish the two-point correlation 
function both at zero and finite temperature. Using this results 
we compute the Casimir force (section \ref{sec:cas}) and the charge density 
of the complex scalar field (section \ref{sec:charged}).  
In section \ref{sec:delta} we consider in details two delta-defects and discover a kind of 
resonance effect in the behavior of the density between the two defects. 
Some remarks and our conclusions are collected in section 
\ref{sec:conclu}.

\section{One defect: reflection-transmission algebra}

It is useful to recall first the algebraic treatment \cite{Mintchev:2003ue} 
of one defect on the real line. 
We take a defect with arbitrary position $y\in \RR$. It divides the real line in two 
domains $D_1=\{x\in \RR\, :\, x<y\}$ and $D_2=\{x\in \RR\, :\, x>y\}$. For simplicity 
we focus on the real massive scalar field $\varphi(t,x)$, satisfying 
\begin{equation}
[\prt_t^2 - \prt_x^2 + m^2] \varphi (t,x) = 0\, , \qquad x\not= y \, .   
\label{eqm}
\end{equation} 
It is well known \cite{A} that all possible dissipationless interactions of $\varphi(t,x)$ with the defect 
$y$ are described by the boundary condition 
\begin{equation}
\left(\begin{array}{cc} \varphi (t,+y) \\ \prt_x \varphi (t,+y)\end{array}\right) = 
\left(\begin{array}{cc} a & b\\ c&d\end{array}\right)
\left(\begin{array}{cc} \varphi (t,-y) \\ \prt_x \varphi (t,-y)\end{array}\right)\, , 
\quad \forall  t\in \RR \, ,   
\label{bc}
\end{equation}
where 
\begin{equation}
ad -bc = 1\, , \qquad  a,...,d \in \RR \, .  
\label{parameters}
\end{equation} 
Following \cite{Mintchev:2004jy, Mintchev:2005rz}, 
one can solve equations (\ref{eqm},\ref{bc}) by introducing the so 
called reflection-transmission algebra with the following structure. 
With each domain $D_i$ one 
associates a ``creation" and ``annihilation" 
operator $\{a^*_i(k),\, a_i(k)\, :\, k\in \RR \}$ which satisfy 
\begin{eqnarray}
{[a_i(k)\, , \,a_j(p)]} &=& 0 = {[a^*_i(k)\,,\,a^*_j (p)]}\, ,\nonumber \\ 
{[a_i(k)\, ,\, a^*_j (p)]} &=&
\left [\delta_{ij}+T_{ij}(k)\right ]2\pi \delta(k-p)+R_{ij}(k) 2\pi \delta(k+p)\, ,
\label{ccr}
\end{eqnarray} 
where 
\begin{equation} 
T(k) =\left(\begin{array}{cc} 0 & T_{12}(k)\\T_{21}(k) & 0\end{array}\right)\, ,
\qquad 
R(k) = \left(\begin{array}{cc} R_{11}(k) & 0\\0 & R_{22}(k)\end{array}\right)\, , 
\label{rtmat}
\end{equation} 
represent the {\it reflection} and {\it transmission coefficients} from the impurity. 
In terms of the parameters $\{a,...,d\}$ one has 
\begin{eqnarray}
R_{11}(k) &=& \frac{bk^2 + \ri (a-d)k + c}{bk^2 +\ri (a+d)k - c}\, 
\e^{-2\ri ky} \ , \qquad 
T_{12}(k)\ =\ \frac{2\ri k}{bk^2 + \ri (a+d)k - c}\, , \qquad
\label{coef1} \\
T_{21}(k) &=& \frac{-2\ri k}{bk^2 - \ri (a+d)k - c} \ , \qquad \qquad
R_{22}(k)\ =\ \frac{bk^2 + \ri (a-d)k + c}{bk^2 - \ri (a+d)k - c}\, 
\e^{-2\ri ky}
\, . \qquad
\label{coef}
\end{eqnarray}
Let us stress that the reflection coefficients (but not the
transmission ones) depend on the position $y$
of the impurity: in the following,  we will also use the `bare' 
coefficients $\wt R_{jj}(k)$ which are just the part independent from
the position $y$:
\begin{eqnarray}
R_{11}(k) &=& \wt R_{11}(k) \,\e^{-2\ri ky} \,, \quad  
R_{22}(k)\ =\ \wt R_{22}(k) \,\e^{-2\ri ky}\, . \qquad
\end{eqnarray}

The associated defect {\it scattering matrix}   
\begin{equation}
S(k) = T(k) + R(k) 
\label{sm1}
\end{equation} 
satisfies unitarity and Hermitian analyticity 
\begin{equation} 
S(k) S(k)^\dag = \II\, ,\qquad S(k)^\dag=S(-k)\, . 
\label{sm2}
\end{equation} 
Now, the solution of (\ref{eqm}) can be written in the form \cite{Mintchev:2004jy} 
\begin{equation}
\varphi (t,x){\Big \vert_{x\in D_i}} \equiv 
\varphi_i (t,x) = \int_{-\infty}^{+\infty} \frac{\rd k}{2\pi \sqrt {2\omega (k)}}
\left[a^*_ i(k) \e^{\ri \omega (k)t-\ri kx} +
a_i (k) \e^{-\ri \omega (k)t+\ri kx}\right ] \,  . 
\label{f1}
\end{equation} 
where $\omega (k) = \sqrt {k^2 + m^2}$. The boundary condition (\ref{bc}) then implies 
that $\{a^*_1(k),\, a_1(k) \}$ and $\{a^*_2(k),\, a_2(k)\}$ are not independent, 
but satisfy the constraints 
\begin{eqnarray}
a_i(k) &=& \sum_{j=1}^2 \Big( R_{ij}(k)\, a_j(-k) + T_{ij}(k)\, a_j(k) \Big) \, ,
\label{c1} \\
a^*_i(k) &=& \sum_{j=1}^2 \Big( a^*_j(-k)\, R_{ji}(-k) + a^*_j(k)\, 
T_{ji}(k)\Big)\,,
\label{c2}
\end{eqnarray}
which are consistent with (\ref{ccr}) because of (\ref{sm2}). 
The algebra $\fA_1$ defined by equations (\ref{ccr}, \ref{c1}, \ref{c2}) 
is known as reflection-transmission (RT) algebra \cite{Mintchev:2003ue}. 
Introducing the matrices 
\begin{equation}
A(k)\equiv \left(\begin{array}{c} a_{1}(k) \\ a_{2}(-k) \end{array}\right)\, ,  
\qquad 
C(k) \equiv \left(\begin{array}{cc} 
R_{11}(k) & T_{12}(k) \\ T_{21}(-k) & R_{22}(-k) \end{array}\right)\, ,  
\label{def1}
\end{equation}
equations (\ref{c1}, \ref{c2}) can be conveniently written in the form 
\begin{equation}
A(k) = C(k)\,A(-k)\, ,\qquad A(k)^\dag = A(-k)^\dag\, C(k)^\dag\, . 
\label{cond1}
\end{equation}
By construction the matrix $C(k)$ satisfies 
\begin{equation}
C(k)\, C(-k)=\II\ , \qquad \ C(k)^\dag = C(-k)^t \, , 
\label{C1}
\end{equation}
where the apex $t$ denotes transposition. 

We end this summary with two observations. In the domain of $\{a,...,d\}$ where 
bound states are absent\cite{Mintchev:2004jy}, the field $\varphi (t,x)$ satisfies 
in addition to (\ref{eqm}, \ref{bc}) the equal-time canonical commutation relations 
\begin{equation}
[\varphi (t,x_1)\, ,\, \varphi (t,x_2)] = 0\, , \qquad
[(\prt_t\varphi )(t,x_1)\, ,\, \varphi (t,x_2)] = -i\delta (x_1-x_2) \, .
\label{initial}
\end{equation} 
We will not consider in this paper the issue of bound states, referring for the details 
about their impact to \cite{Mintchev:2004jy}. 

We stress in conclusion that the above construction is purely algebraic and applies for any 
representation of the RT algebra $\fA_1$. This fact allows to investigate the physical effects 
associated with nonequivalent representations and in particular, 
to introduce \cite{Mintchev:2004jy} the concept of 
temperature, which is relevant for the applications in section 5.

\section{The RT algebra of two defects\label{sec:2defect}}

We show now how the RT algebraic framework allows to compute in a 
very systematic and straightforward way the global reflection and 
transmission coefficients when there are two defects. We confirm the 
results of \cite{SCH}, where these coefficients where computed 
through the calculation of Green functions.

We consider two defects located at $y_{1}<y_{2}$. At each of them 
\begin{equation}
\left(\begin{array}{cc} \varphi (t,+y_\alpha ) \\ \prt_x \varphi (t,+y_\alpha )\end{array}\right) = 
\left(\begin{array}{cc} a_\alpha & b_\alpha\\ c_\alpha&d_\alpha \end{array}\right)
\left(\begin{array}{cc} \varphi (t,-y_\alpha) \\ \prt_x \varphi (t,-y_\alpha)\end{array}\right)\, , 
\quad \alpha = 1,2 \, .  
\label{bc2}
\end{equation} 
Let $R^{[\alpha]}(k)$ and $T^{[\alpha]}(k)$ be the reflection and transition matrices 
defined by (\ref{coef}) in terms of the parameters $\{a_\alpha,...,d_\alpha\}$. 
According to (\ref{def1}) we set 
\begin{equation} 
C^{[1]}(k) \equiv \left(\begin{array}{cc} 
R^{[1]}_{11}(k) & T^{[1]}_{12}(k) \\ 
T^{[1]}_{21}(-k) & R^{[1]}_{22}(-k) \end{array}\right)\, ,  \qquad 
C^{[2]}(k) \equiv \left(\begin{array}{cc} 
R^{[2]}_{22}(k) & T^{[2]}_{23}(k) \\ 
T^{[2]}_{32}(-k) & R^{[2]}_{33}(-k) \end{array}\right)\, ,  
\label{def2}
\end{equation}
which satisfy by construction 
\begin{equation}
C^{[\alpha ]}(k)\, C^{[\alpha ]}(-k)=\II\ , \qquad 
C^{[\alpha ]}(k)^\dag = C^{[\alpha ]}(-k)^t \, , \qquad \alpha = 1,2\, . 
\label{cunit}
\end{equation}
The two defects separate now the real line in three 
domains 
\begin{equation} 
D_1=\{x\in \RR\, :\, x<y_1\}\, , \quad 
D_2=\{x\in \RR\, :\, y_1<x<y_2\}\, , \quad 
D_3=\{x\in \RR\, :\, y_2<x \}\, . 
\label{domains1}
\end{equation}
{}Following the previous section, we associate with each of them the 
operators $\{a^*_i(k),\, a_i(k)\}$ and define the field $\varphi (t,x)$ by 
(\ref{f1}), where now $i=1,2,3$. Since $\varphi (t,x)$ must satisfy (\ref{bc2}), we impose 
\begin{equation}
A^{[\alpha ]}(k) = C^{[\alpha ]}(k)\, A^{[\alpha ]}(-k)\, ,\qquad 
A^{[\alpha ]}(k)^\dag = A^{[\alpha ]}(-k)^\dag\, C^{[\alpha ]}(k)^\dag\, ,
\qquad \alpha = 1,2\, ,
\label{cond2}
\end{equation}
where 
\begin{equation}
A^{[\alpha ]}(k) \equiv \left(\begin{array}{c} a_{\alpha }(k) \\ 
a_{\alpha +1}(-k) \end{array}\right)\, ,  
\qquad \alpha = 1,2\, . 
\end{equation}
The operators $\{a^*_i(k),\, a_i(k) \}$ describe the propagation of 
$\varphi (t,x)$ in $D_i$. Only $\{a^*_1(k),\, a_1(k) \}$ and $\{a^*_3(k),\, a_3(k) \}$ 
give origin to asymptotic states. Therefore it is natural to expect that 
$\{a^*_2(k),\, a_2(k) \}$ can be expressed in terms of $\{a^*_1(k),\, a_1(k) \}$ and 
$\{a^*_3(k),\, a_3(k) \}$. Indeed, from (\ref{cond2}) it follows that 
\begin{eqnarray}
C^{[2]}_{11}(k) a_2(-k)&=&a_2(k)- C^{[2]}_{12}(k) a_3(k)\, ,\nonu 
C^{[1]}_{22}(-k) a_2(-k)&=&a_2(k)- C^{[1]}_{21}(-k) a_1(k)\, . 
\end{eqnarray}
Eliminating $a_2(-k)$ from these two equations, one gets 
\begin{equation}
\left [C^{[2]}_{11}(k)-C^{[1]}_{22}(-k)\right ] a_{2}(k) = 
C^{[2]}_{11}(k)\,C^{[1]}_{21}(-k)\,a_{1}(k)-C^{[1]}_{22}(-k)\,C^{[2]}_{12}(k)\, a_{3}(k)\, . 
\label{eq:a2-a13}
\end{equation}
Remark that because of the factors $\e^{2\ri ky_{2}}$ and $\e^{2\ri ky_{1}}$ 
appearing in $R^{[2]}_{22}(k)$ and $R^{[1]}_{22}(k)$ respectively, 
the coefficient $C^{[2]}_{11}(k)-C^{[1]}_{22}(-k)$ never vanishes for generic $y_1, y_2$  and 
(\ref{eq:a2-a13}) uniquely determines $a_2(k)$ in terms of $a_1(k)$ and $a_3(k)$. 
One has 
\begin{equation}
a_2(k) = \frac{R^{[2]}_{22}(k) T^{[1]}_{21}(-k) a_1(k) - R^{[1]}_{22}(k) T^{[2]}_{23}(k) a_3(k)}
{R^{[2]}_{22}(k) - R^{[1]}_{22}(k)}\, , 
\label{a2}
\end{equation}
Analogously, $a^*_2(k)$ is given in terms of $a^*_1(k)$ and $a^*_3(k)$, 
\begin{equation}
a^*_{2}(k) = \frac{R^{[2]}_{22}(-k) T^{[1]}_{12}(-k) a^*_1(k) - R^{[1]}_{22}(-k) T^{[2]}_{32}(k) a^*_3(k)}
{R^{[2]}_{22}(-k) - R^{[1]}_{22}(-k)}\, ,
\label{a*2}
\end{equation} 
We conclude therefore that in order to define the 
RT algebra for two impurities $\fA_2$, it is 
enough to fix the commutation relations between $\{a^*_1(k),\, a_1(k) \}$ and 
$\{a^*_3(k),\, a_3(k) \}$. For this purpose we have to determine the global  
reflection and transmission matrices 
\begin{equation} 
\cT(k) =\left(\begin{array}{cc} 0 & \cT_{13}(k)\\ \cT_{31}(k) & 0\end{array}\right)\, ,
\qquad 
\cR(k) = \left(\begin{array}{cc} \cR_{11}(k) & 0\\0 & \cR_{33}(k)\end{array}\right)
\label{rtmat2}
\end{equation} 
in terms of the local ones $R^{[\alpha ]}$ and $T^{[\alpha ]}$. 
According to (\ref{c1}), $\cT (k)$ and $\cR (k)$ can be read from the relations between 
$a_1(k)$ and $a_3(k)$, which can be derived as follows. 
Plugging the expression (\ref{eq:a2-a13}) into (\ref{cond2}), 
we get
\begin{equation}
a_{1}(k) =
C^{[1]}_{22}(-k)^{-1}\,C^{[1]}_{11}(k)\,
\frac{C^{[1]}_{22}(-k)-C^{[2]}_{11}(k)}{
1-C^{[1]}_{22}(k)\,C^{[2]}_{11}(k)}\,a_{1}(-k) +
\frac{C^{[1]}_{12}(k)\,C^{[2]}_{12}(k)}{ 
1-C^{[1]}_{22}(k)\,C^{[2]}_{11}(k)}\,a_{3}(k) \, . 
\label{r1}
\end{equation} 
By means of the identities 
$$
C^{[1]}_{22}(-k)^{-1} = C^{[1]}_{22}(k) + 
C^{[1]}_{22}(-k)^{-1}\,C^{[1]}_{12}(k)\,C^{[1]}_{21}(-k)\, , 
$$
$$
C^{[1]}_{12}(-k)\,C^{[1]}_{22}(-k)^{-1} =  -C^{[1]}_{11}(k)^{-1}\,C^{[1]}_{12}(k)\, ,
$$
{}following from (\ref{cunit}), equation (\ref{r1}) can be rewritten in the form 
\begin{equation}
a_{1}(k) = \left [C^{[1]}_{11}(k)
+\frac{C^{[1]}_{12}(k)\,C^{[1]}_{21}(k)\,C^{[2]}_{11}(k)}{
1-C^{[1]}_{22}(k)\,C^{[2]}_{11}(k) }\right ] a_{1}(-k) +
\frac{C^{[1]}_{12}(k)\,C^{[2]}_{12}(k)}{
1-C^{[1]}_{22}(k)\,C^{[2]}_{11}(k) }\,a_{3}(k) \, . 
\label{r2}
\end{equation} 
The same type of calculations provides
\begin{equation}
a_{3}(k) = \left [C^{[2]}_{22}(-k)
+\frac{C^{[2]}_{21}(-k)\,C^{[2]}_{12}(-k)\,C^{[1]}_{22}(-k)}{
1-C^{[1]}_{22}(-k)\,C^{[2]}_{11}(-k)}\right ] a_{3}(-k) +
\frac{C^{[2]}_{21}(-k)\,C^{[1]}_{21}(-k)}{
1-C^{[1]}_{22}(-k)\,C^{[2]}_{11}(-k)} \,a_{1}(k) \, . 
\label{r3}
\end{equation}
Comparing (\ref{r2}, \ref{r3}) with (\ref{c1}), one infers 
\begin{eqnarray}
\cR_{11}(k) &=&  C^{[1]}_{11}(k)
+\frac{C^{[1]}_{12}(k)\,C^{[1]}_{21}(k)\,C^{[2]}_{11}(k)}{
1-C^{[1]}_{22}(k)\,C^{[2]}_{11}(k) }\ :=\ \wt\cR_{11}(k)\,\e^{-2\ri ky_{1}}
\label{r11}\\
\wt\cR_{11}(k) &=& \wt R^{[1]}_{11}(k)
+\frac{T^{[1]}_{12}(k)\,T^{[1]}_{21}(-k)\,\wt R^{[2]}_{22}(k)\,\e^{2\ri ky_{12}}}{
1-\wt R^{[1]}_{22}(-k)\,\wt R^{[2]}_{22}(k)\,\e^{2\ri ky_{12}} }\, , 
\label{eq:r11tilde}
\end{eqnarray}
\begin{eqnarray}
\cR_{33}(k) &=&
C^{[2]}_{22}(-k)
+\frac{C^{[2]}_{21}(-k)\,C^{[2]}_{12}(-k)\,C^{[1]}_{22}(-k)}{
1-C^{[1]}_{22}(-k)\,C^{[2]}_{11}(-k)}\ :=\ \wt\cR_{33}(k)\,\e^{-2\ri ky_{2}}
\label{r33}\\
\wt\cR_{33}(k) &=& \wt R^{[2]}_{33}(k)
+\frac{T^{[2]}_{32}(k)\,T^{[2]}_{23}(-k)\,
\wt R^{[1]}_{22}(k)\,\e^{-2\ri ky_{12}}}{
1-\wt R^{[1]}_{22}(k)\,\wt R^{[2]}_{22}(-k)\,\e^{-2\ri ky_{12}}}\, , 
\label{eq:r33tilde}
\end{eqnarray}
\begin{eqnarray}
\cT_{13}(k) = 
\frac{C^{[1]}_{12}(k)\,C^{[2]}_{12}(k)}{
1-C^{[1]}_{22}(k)\,C^{[2]}_{11}(k) }= 
\frac{T^{[1]}_{12}(k)\,T^{[2]}_{23}(k)}{ 
1-\wt R^{[1]}_{22}(-k)\,\wt R^{[2]}_{22}(k)\,\e^{2\ri ky_{12}}} \, ,
\label{t13}
\end{eqnarray}
\begin{eqnarray}
\cT_{31}(k)= \frac{C^{[2]}_{21}(-k)\,C^{[1]}_{21}(-k)}{
1-C^{[1]}_{22}(-k)\,C^{[2]}_{11}(-k)} = 
\frac{T^{[2]}_{32}(k)\,T^{[1]}_{21}(k)}{
1-\wt R^{[1]}_{22}(k)\,\wt R^{[2]}_{22}(-k)\,\e^{-2\ri ky_{12}}}\, , 
\label{t31}
\end{eqnarray} 
which express the global two-impurity reflection and transmission coefficients in 
terms of the local ones. As for the one-impurity case, we have 
introduced `bare' reflection coefficients which, as the transmission 
coefficients, depend only on the distance between the 
two impurities, $y_{12}=y_{1}-y_{2}$. 

Remark that when one of the two 
defect is `trivial', i.e. $R_{11}(k)=R_{22}(k)=0$ and 
$T_{12}(k)=T_{21}(k)=1$, the global coefficients $\cR$ and $\cT$ 
become identical to the ones of the remaining defect.

The relative $S$-matrix is 
\begin{equation}
S(k) = \cT(k) + \cR(k) \, . 
\label{sm3}
\end{equation}
Using the expressions (\ref{r33}-\ref{t31}) and the properties of $R^{[\alpha ]}(k)$ 
and $T^{[\alpha ]}(k)$ one can verify that (\ref{sm3}) satisfies both 
unitarity and Hermitian analyticity (\ref{sm2}), which are fundamental for the construction 
of the RT algebra $\fA_2$ below. 

Equations (\ref{r33}-\ref{t31}) describe in a compact way all (possible infinite) processes 
of reflection and transmission from the two defects, relating an incoming to an outgoing 
wave. Accordingly, we impose on the $\fA_2$-generators  
\begin{eqnarray}
{[a_i(k)\,,\,a_j(p)]} &=& 0 = {[a^*_i(k)\,,\,a^*_j(p)]}\, , \nonumber 
\\
{[a_i(k)\,,\,a^*_j(p)]} &=&
\left [\delta_{ij}+\cT_{ij}(k)\right ] 2\pi \delta(k-p)+\cR_{ij}(k)2\pi \delta(k+p)\, , 
\label{ccr1}
\end{eqnarray}
where $i,j$ take {\it only} the values 1 and 3. The exchange relations involving 
$\{a^*_2(k),\, a_2(k) \}$ are determined by (\ref{ccr1}) and the relations 
(\ref{a2}, \ref{a*2}). 

Summarizing, the RT algebra $\fA_2$ for two defects 
has the following simple structure. The elements $\{a^*_2(k),\, a_2(k) \}$, associated with  
the field evolution between the defects, are linear combinations of $\{a^*_1(k),\, a_1(k) \}$ 
and $\{a^*_3(k),\, a_3(k) \}$, which generate the asymptotic states and satisfy 
quadratic exchange relations (\ref{ccr1}). The latter have the same form as in 
the one-impurity case (\ref{ccr}), but involve the global two-impurity 
reflection and transmission matrices $\cR(k)$ and $\cT(k)$, which are uniquely 
determined by the single defects via (\ref{r11}-\ref{t31}). We will show in the next section 
that this result has a straightforward generalization to an arbitrary number 
$n>2$ 
of defects. 

\section{General case of several defects\label{sec:Ndef}}

We consider here ${n}$ defects on $\RR$ with coordinates $y_1< \cdots 
<y_{n}$. They 
define the ${n}+1$ domains 
\begin{equation} 
D_1=\{x\in \RR\, :\, x<y_1\}\, , ... ,\, 
D_i=\{x\in \RR\, :\, y_{i-1}<x<y_i\}\, , ... ,\, 
D_{{n}+1}=\{x\in \RR\, :\, y_{n}<x\}\, . 
\label{domains2}
\end{equation}
The relative RT algebra $\fA_{n}$ is therefore generated by $\{a^*_i(k),\, a_i(k)\, :\, i=1,...,{n}+1\}$. 
Extrapolating from the previous section, we expect 
that all $\{a^*_i(k),\, a_i(k)\, :\, i=2,...,{n}\}$ can be expressed in terms 
of $\{a^*_1(k),\, a_1(k)\}$ and $\{a^*_{{n}+1}(k),\, a_{{n}+1}(k)\}$, which generate in turn 
the asymptotic states. Let us now verify this scenario and determine the 
algebra $\fA_{n}$. Setting  
\begin{equation} 
C^{[\alpha ]}(k) \equiv \left(\begin{array}{cc} 
R^{[\alpha ]}_{\alpha \alpha}(k) & T^{[\alpha ]}_{\alpha (\alpha+1)}(k) \\ 
T^{[\alpha ]}_{(\alpha+1) \alpha }(-k) & R^{[\alpha]}_{(\alpha+1)(\alpha +1)}(-k) \end{array}\right)\, , 
\qquad \alpha = 1,...,{n}\, , 
\label{defn}
\end{equation} 
we infer from the boundary condition (\ref{bc2}) that (\ref{cond2}, \ref{a2}) hold now for any 
$\alpha = 1,...,{n}$. 

Indeed, applying recursively the technics used in the case of two 
defects, one deduces that 
\begin{equation}
A^{[j,l]}(k)=\left(\begin{array}{c} a_{j}(k) \\ a_{l}(-k) \end{array}\right)\, ,
\qquad j,k=1,\ldots,{n}+1\, , 
\end{equation}
obeys the relations
\begin{eqnarray}
A^{[j,l]}(k) &=& C^{[j\ldots l]}(k)\,A^{[j,l]}(-k)\, ,\\
C^{[j\ldots l]}(k)^\dag &=& C^{[j\ldots l]}(-k)^t
\mb{and}C^{[j\ldots l]}(k)\,C^{[j\ldots l]}(-k)=\II \, ,
\quad 1\leq j<l\leq {n}+1 \, , 
\end{eqnarray}
where $C^{[j\ldots l]}(k)$ is some matrix built on the $C^{[\alpha]}(k)$ matrices,
$\alpha=j,\ldots,l$.
Starting from 
\begin{eqnarray}
a_{j}(k) &=& C^{[1\ldots j]}_{21}(k)\,a_{1}(k) + 
C^{[1\ldots j]}_{22}(k)\,a_{j}(-k)\, , 
\\
a_{j}(k) &=& C^{[j\ldots {n}+1]}_{11}(k)\,a_{j}(-k) + 
C^{[j\ldots {n}+1]}_{12}(k)\,a_{{n}+1}(k)\, ,  
\end{eqnarray}
one gets
\begin{equation}
\Big(C_{11}^{[j\ldots {n}+1]}(k)-C_{22}^{[1\ldots j]}(k)\Big)\,a_{j}(k) = 
C_{21}^{[1\ldots j]}(k)\,C_{11}^{[j\ldots {n}+1]}(k)\,a_{1}(k)
-C_{22}^{[1\ldots j]}(k)\,C_{12}^{[j\ldots {n}+1]}(k)\,a_{{n}+1}(k).
\end{equation}
Therefore, the expressions found in the case of two defects also apply
here, with the rules
\begin{eqnarray*}
    C^{[1]}(k)\,\to\,C^{[1\ldots j]}(k) \ ;\ C^{[2]}(k)\,\to\,C^{[j\ldots
{n}+1]}(k)\, , \\
a_{1}(k)\,\to\,a_{1}(k)\ ;\ a_{2}(k)\,\to\,a_{j}(k)
\ ;\ a_{3}(k)\,\to\,a_{{n}+1}(k)\, .
\end{eqnarray*}
Hence, it remains to compute the matrices $C^{[j\ldots l]}(k)$ from the
matrices $C^{[\alpha]}(k)$. We give here a recursion formula for them.
We consider the relations
\begin{eqnarray}
A^{[1]}(k) &=& C^{[1]}(k)\,A^{[1]}(-k)\, ,\\
C^{[1]}(k)^\dag &=& C^{[1]}(-k)^t 
\mb{and}C^{[1]}(k)\,C^{[1]}(-k)=\II\ , \\
A^{[2,{n}+1]}(k) &=& C^{[2\ldots {n}+1]}(k)\,A^{[2,{n}+1]}(-k)\, ,\\
C^{[2\ldots {n}+1]}(k)^\dag &=& C^{[2\ldots {n}+1]}(-k)^t
\mb{and}C^{[2,{n}+1]}(k)\,C^{[2\ldots {n}+1]}(-k)=\II \, , 
\nonumber
\end{eqnarray}
which amounts to take $j=2$ in the previous calculations.
This leads to
\begin{eqnarray}
A^{[1,{n}+1]}(k) &=& C^{[1\ldots {n}+1]}(k)\,A^{[1,{n}+1]}(-k)\, , 
\end{eqnarray}
or in components 
\begin{eqnarray}
a_{1}(k) &=& C^{[1\ldots {n}+1]}_{11}(k)\,a_{1}(-k) + 
C^{[1\ldots {n}+1]}_{12}(k)\,a_{{n}+1}(-k)\, ,\\
a_{{n}+1}(k) &=& C^{[1\ldots {n}+1]}_{21}(k)\,a_{1}(-k) + 
C^{[1\ldots {n}+1]}_{22}(k)\,a_{{n}+1}(-k)\, ,
\end{eqnarray}
where 
\begin{eqnarray}
C^{[1\ldots {n}+1]}_{11}(k) &=& C^{[1]}_{11}(k)
+\frac{C^{[1]}_{12}(k)\,C^{[1]}_{21}(k)\,C^{[2\ldots {n}+1]}_{22}(k)}{
1-C^{[1]}_{22}(k)\,C^{[2\ldots {n}+1]}_{11}(k) }\, , 
\\
C^{[1\ldots {n}+1]}_{12}(k) &=& 
\frac{C^{[1]}_{12}(k)\,C^{[2\ldots {n}+1]}_{12}(k)}{
1-C^{[1]}_{22}(k)\,C^{[2\ldots {n}+1]}_{11}(k) }\, ,
\\
C^{[1\ldots {n}+1]}_{21}(k) &=& 
\frac{C^{[2\ldots {n}+1]}_{21}(k)\,C^{[1]}_{21}(k)}{
1-C^{[1]}_{22}(k)\,C^{[2\ldots {n}+1]}_{11}(k)}\, ,
\\
C^{[1\ldots {n}+1]}_{22}(k) &=& C^{[2\ldots {n}+1]}_{22}(k)
+\frac{C^{[2\ldots {n}+1]}_{21}(k)\,C^{[2\ldots {n}+1]}_{12}(k)\,C^{[1]}_{22}(k)}{
1-C^{[1]}_{22}(k)\,C^{[2\ldots {n}+1]}_{11}(-k)}\, .
\end{eqnarray}
Hence, an iterative use of these formulae, starting from the `local' 
coefficients $C^{[\alpha]}(k)$, allows to compute all the $C^{[j\ldots l]}(k)$.

Alternative recursion formulas can be obtained particularizing the 
last defect instead of the first one. One gets

\begin{eqnarray}
C^{[1\ldots {n}+1]}_{11}(k) &=& C^{[1\ldots {n}]}_{11}(k)
+\frac{C^{[1\ldots {n}]}_{12}(k)\,C^{[1\ldots {n}]}_{21}(k)\,C^{[{n}+1]}_{22}(k)}{
1-C^{[1\ldots {n}]}_{22}(k)\,C^{[{n}+1]}_{11}(k) }\, , 
\\
C^{[1\ldots {n}+1]}_{12}(k) &=& 
\frac{C^{[1\ldots {n}]}_{12}(k)\,C^{[{n}+1]}_{12}(k)}{
1-C^{[1\ldots {n}]}_{22}(k)\,C^{[{n}+1]}_{11}(k) }\, ,
\\
C^{[1\ldots {n}+1]}_{21}(k) &=& 
\frac{C^{[{n}+1]}_{21}(k)\,C^{[1\ldots {n}]}_{21}(k)}{
1-C^{[1\ldots {n}]}_{22}(k)\,C^{[{n}+1]}_{11}(k)}\, ,
\\
C^{[1\ldots {n}+1]}_{22}(k) &=& C^{[{n}+1]}_{22}(k)
+\frac{C^{[{n}+1]}_{21}(k)\,C^{[{n}+1]}_{12}(k)\,C^{[1\ldots {n}]}_{22}(k)}{
1-C^{[1\ldots {n}]}_{22}(k)\,C^{[{n}+1]}_{11}(-k)}\, .
\end{eqnarray}

Note also that one can deduce physical properties directly from
the recursion formulae.
In particular, from unitarity and hermiticity relations of
$C^{[1]}(k)$ and $C^{[2\ldots {n}+1]}(k)$, it is easy to check that these
relations are also obeyed by $C^{[1\ldots {n}+1]}(k)$:
\begin{eqnarray}
C^{[1\ldots {n}+1]}(k)\,C^{[1\ldots {n}+1]}(-k)=\II
\mb{and}C^{[1\ldots {n}+1]}(k)^\dag = C^{[1\ldots {n}+1]}(-k)^t \, .
\end{eqnarray}
In the same way, one proves recursively that the transmission 
coefficients $C^{[1\ldots {n}+1]}_{12}(k)$ and $C^{[1\ldots {n}+1]}_{21}(k)$ 
depend only on the distances between the different defects, while the 
reflection coefficients obey
$$
C^{[1\ldots {n}+1]}_{11}(k)=\wt C^{[1\ldots {n}+1]}_{11}(k)\,\e^{-\ri 
k\,y_{1}} \mb{and} 
C^{[1\ldots {n}+1]}_{22}(k)=\wt C^{[1\ldots {n}+1]}_{22}(k)\,\e^{\ri 
k\,y_{{n}+1}}\, ,
$$
where the `bare' coefficients $\wt C^{[1\ldots {n}+1]}_{jj}(k)$ 
depend only on the distances between the different defects.

\section{Quantum fields with defects\label{sec:fieldtheo}} 

This section is devoted to some aspects of quantum field theory with defects on the line. 
In order to keep the discussion as simple as possible we consider one scalar field 
and ${n}=2$ defects. The results however can be easily extended to more general 
cases using the results of section \ref{sec:Ndef}. 

\subsection{Correlation functions} 

In the Fock representation $\cF(\fA_2)$ of $\fA_2$ the basic 
correlation functions are \cite{Mintchev:2004jy}
\begin{equation}
\langle a_i(k )a^*_j(p)\rangle = 
2\pi \left\{\left[\delta_{ij} + \cT_{ij}(k)\right] \delta (k-p) 
+ \cR_{ij}(k) \delta (k+p)  \right \}\, , \quad 
\langle a^*_i(k )a_j(p)\rangle = 0\, , 
\label{fock1}
\end{equation}
where $i,j=1,3$ and $\cR$ and $\cT$ are given by 
(\ref{r11}-\ref{t31}), together with
\begin{eqnarray}
\cT_{11}(k) &=& \cT_{33}(k) = 0\, , \qquad \cR_{13}(k) = \cR_{31}(k) =0\, . 
\label{cf1}
\end{eqnarray} 
Because of (\ref{a2}, \ref{a*2}), the correlators involving 
$a^*_2(k)$ and $a_2(k)$ can be expressed in terms of (\ref{fock1}). 
Now, using (\ref{f1}) one has 
\begin{equation} 
\langle \varphi (t_1,x_1 ) \varphi (t_2,x_2 )\rangle {\Big \vert_{x_1\in D_i,\, x_2\in D_j}} = 
\langle \varphi_i (t_1,x_1 ) \varphi_j (t_2,x_2 )\rangle 
= \overline{\langle \varphi_j (t_2,x_2 ) \varphi_i (t_1,x_1 )\rangle }\, ,  
\label{fock2}
\end{equation} 
which can be computed for any $i,j=1,2,3$ and have the following general form 
\begin{equation}
\langle \varphi_i (t_1,x_1 ) \varphi_j (t_2,x_2 )\rangle = 
\int_{-\infty}^{+\infty} \frac{\rd k}{4\pi \omega (k)} \e^{-\ri \omega(k) t_{12}}
\left \{[\delta_{ij}+\cT_{ij}(k)]\,\e^{\ri kx_{12}} + \cR_{ij}(k)\,
\e^{\ri k\tx_{12}}\right \}\, , 
\label{cf0}
\end{equation} 
where $t_{12}=t_1-t_2$, $x_{12}=x_1-x_2$ and $\tx_{12}=x_1+x_2$. We recall that the dispersion 
relation is 
\begin{equation}
\omega(k) = \sqrt {k^2 + m^2} \, . 
\label{dispersion}
\end{equation} 

{}For the correlators involving $i$ or $j=2$,
we use the expressions (\ref{a2}, \ref{a*2}) 
and after some algebra we find 
\begin{eqnarray}
\cT_{22}(k)&=& \frac{\wt R^{[1]}_{22}(-k)\wt R^{[2]}_{22}(k)\,\e^{2\ri ky_{12}} }
{1-\wt R^{[1]}_{22}(-k)\wt R^{[2]}_{22}(k)\,\e^{2\ri ky_{12}} } + 
\frac{\wt R^{[1]}_{22}(k)\wt R^{[2]}_{22}(-k)\,\e^{-2\ri ky_{12}} }
{1-\wt R^{[1]}_{22}(k)\wt R^{[2]}_{22}(-k)\,\e^{-2\ri ky_{12}}} \, , \\
\cR_{22}(k)&=&
\frac{\wt R^{[2]}_{22}(k)\,\e^{-2\ri ky_{2}}}
{1-\wt R^{[1]}_{22}(-k)\wt R^{[2]}_{22}(k)\,\e^{2\ri ky_{12}} } + 
\frac{\wt R^{[1]}_{22}(k)\,\e^{-2\ri ky_{1}}}
{1-\wt R^{[1]}_{22}(k)\wt R^{[2]}_{22}(-k)\,\e^{-2\ri ky_{12}} } \, ,
\label{cf2}
\end{eqnarray}
\begin{eqnarray}
\cT_{12}(k)=
\frac{T^{[1]}_{12}(k)}{1-\wt R^{[1]}_{22}(-k)\wt R^{[2]}_{22}(k)\,\e^{2\ri ky_{12}} }\, , \qquad
\cR_{12}(k)= 
\frac{T^{[1]}_{12}(k)\wt R^{[2]}_{22}(k)\,\e^{-2\ri ky_{2}}}
{1-\wt R^{[1]}_{22}(-k)\wt R^{[2]}_{22}(k)\,\e^{2\ri ky_{12}}}\, ,
\label{cf4}
\end{eqnarray}
\begin{eqnarray}
\cT_{23}(k)=\frac{T^{[2]}_{23}(k)}{1- \wt R^{[1]}_{22}(-k) 
\wt R^{[2]}_{22}(k)\,\e^{2\ri ky_{12}}}\, , \qquad 
\cR_{23}(k)=\frac{T^{[2]}_{23}(-k) \wt R^{[1]}_{22}(k)\,\e^{-2\ri ky_{1}}}
{1- \wt R^{[1]}_{22}(k)\wt R^{[2]}_{22}(-k)\,\e^{-2\ri ky_{12}}}\, ,
\label{cf6}
\end{eqnarray}
where we have used `bare' coefficients to make the $y_{j}$ dependence 
explicit.

The remaining coefficients are computed thanks to (\ref{fock2}), that is 
to say
$$
\cR_{ij}(k)^\dag = \cR_{ji}(-k) \mb{and} \cT_{ij}(k)^\dag = 
\cT_{ji}(k)\,,\quad i,j=1,2,3.
$$

\subsection{Casimir force\label{sec:cas}} 

The energy density of the field $\varphi (t,x)$ reads 
\begin{equation}
\theta (t,x)  =  \frac{1}{2}\left \{ (\partial_t\varphi)^2 -  
\frac{1}{2}\left [ \varphi(\partial^2_x\varphi) + (\partial^2_x\varphi) \varphi \right ] + 
m^2 \varphi^2 \right \}(t,x)\, . 
\label{cas1}
\end{equation}
One can compute the expectation value $ \langle \theta (t,x,i) \rangle $ via 
point-splitting from the two-point function (\ref{cf0}). Like in the standard Casimir effect, 
this expectation value diverges at coinciding points. This fact should not be surprising because 
it is the force, which is actually the physical observable in this context. It is given by  
\begin{equation}
\cF_i(x ) = \frac{\rd}{\rd x} \langle \theta (t,x) \rangle {\Big \vert_{x\in D_i}}=  
\ri \int_{-\infty}^{+\infty} \frac{\rd k}{2\pi}\, k\, \omega (k)\, 
\cR_{ii}(k) \e^{2\ri k x}\, , \qquad i=1,2,3 \, , 
\label{for-cas}
\end{equation} 
which determines a well-defined {\it distribution} in $x$. In general, $\cF_i(x )$ may be singular 
at $x=y_i$. We will illustrate this feature below on the example of two delta defects.

\subsection{Finite temperature} 

In order to investigate the Casimir force at finite (inverse) 
temperature $\beta$, we introduce the so called Gibbs representation 
$\cG_\beta (\fA_2)$ of $\fA_2$. In contrast to the Fock vacuum, 
the fundamental state of $\cG_\beta (\fA_2)$ is not annihilated by 
$a_i(k)$ and represents an appropriate idealization of a thermal 
bath which keeps the system in equilibrium. In $\cG_\beta (\fA_2)$
one has \cite{Mintchev:2004jy}
\begin{equation}
\langle a_i(k )a^*_j(p)\rangle_\beta = \frac{1}{1-\e^{-\beta \omega (k)}}\,  
2\pi \left\{\left[\delta_{ij} + \cT_{ij}(k)\right] \delta (k-p) 
+ \cR_{ij}(k) \delta (k+p)  \right \}\, , 
\label{gibbs1}
\end{equation}
\begin{equation}
\langle a_i^*(k )a_j(p)\rangle_\beta = \frac{\e^{-\beta \omega (k)}}{1-\e^{-\beta \omega (k)}}\, 
2\pi \left\{ \left[ \delta_{ij} + \cT_{ji}(k)\right] \delta (k-p) 
+ \cR_{ji}(-k) \delta (k+p)  \right \} \, , 
\label{gibbs2}
\end{equation}
where $i,j = 1,3$. It is easy to show at this point that the thermal correlation functions 
$\langle \varphi_i (t_1,x_1 ) \varphi_j (t_2,x_2 )\rangle_\beta $ are 
obtained by substitution 
\begin{equation} 
\e^{-\ri \omega(k)t_{12}} \longmapsto  
\frac{\e^{-\beta \omega (k)+\ri \omega(k)t_{12}} + \e^{-\ri \omega(k)t_{12}}}{1-\e^{-\beta \omega (k)}} 
\label{gibbs3}
\end{equation}
in the integrand of the respective zero-temperature expressions (\ref{cf0}). 
In this way one now gets for the force 
\begin{equation} 
\cF_i(x, \beta )  = \ri\int_{-\infty}^{\infty}\frac{\rd k}{2\pi}\, 
\frac{k\,\omega (k)}{1-\e^{-\beta \omega (k)}} 
\, \cR_{ii}(k)\, \e^{2\ri k x } \, ,\quad
 i=1,2,3\, , 
 \label{tcforce}
\end{equation} 
showing an explicit temperature dependence. 

\subsection{Charge density\label{sec:charged}} 

In this section we consider a complex scalar field 
\begin{equation}
\varphi (t,x) = \frac{1}{\sqrt 2} \left [\varphi^{(1)} (t,x) + 
i \varphi^{(2)} (t,x) \right ] \, , 
\end{equation}
where both $\varphi^{(1)}$ and $\varphi^{(2)}$ are two copies of the 
scalar field with the same two defects constructed in section 3 above. 
The theory admits now a $U(1)$ symmetry leading to the conservation of the  
current 
\begin{equation} 
j_\mu (t,x) = -i \left [:(\prt_\mu \varphi^\ast )\varphi :(t,x) - 
:\varphi^\ast (\prt_\mu \varphi ) :(t,x) \right ] \, ,  
\label{current} 
\end{equation} 
where $:\cdots :$ stands for the normal product in the RT algebra $\fA_2$. 
We are interested in the charge density 
\begin{equation}
\varrho(x,\beta ) = \langle j_0(t,x) \rangle_\beta   
\label{cd1}
\end{equation}
of the system in the Gibbs state, defined in the previous subsection. 
Using the thermal two-point function of $\varphi $, one obtains 
\begin{equation}
\varrho (x,\beta){\Big \vert_{x\in D_i}} \equiv \varrho_i(x,\beta)  = 
\int_{-\infty}^{\infty} \frac{dk}{\pi}
\frac{\e^{-\beta \omega(k)}} 
{ 1-\e^{-\beta \omega(k)}} \left (1+\cR_{ii}(k) \e^{2 i k x}\right ) \, .  
\label{cd2}
\end{equation} 
The density (\ref{cd2}) vanishes in the limit of 
zero temperature ($\beta \to \infty$): the first term in the integrand gives 
the density at finite temperature, whereas the second one generates 
the correction due to the defects. 

\section{Example: two delta impurities\label{sec:delta}}

We consider the case of two defects defined by the boundary 
conditions
\begin{equation}
\begin{array}{l}
 \varphi (t,+y_{\alpha}) = \varphi (t,-y_{\alpha})\, , \\[1.2ex] 
\prt_x \varphi (t,+y_{\alpha}) - \prt_x \varphi (t,-y_{\alpha})
=\mu_{\alpha}\,\varphi (t,-y_{\alpha})\, , 
\end{array}
\quad \alpha=1,2\,,
\quad \forall  t\in \RR \, .   
\end{equation}
The local reflection and transmission coefficients have the form
\begin{equation}
\wt R^{[\alpha]}_{11}(k) =\wt R^{[\alpha]}_{22}(-k) = 
\frac{-i\mu_{\alpha}}{k+i\mu_{\alpha}} 
\mb{and}
T^{[\alpha]}_{12}(k) = T^{[\alpha]}_{21}(-k) = 
\frac{k}{k+i\mu_{\alpha}} \,,\quad \alpha=1,2\,.\quad
\end{equation}
To avoid the presence of bound states we will assume that the product
$\mu_{1}\,\mu_{2}$ is strictly positive. In the computation of the Casimir force 
we will also suppose for simplicity that $m=0$.

\subsection{Force on a test particle between the defects\label{sec:CasMid}}

\begin{figure}[htb]
\begin{center}
\epsfig{file=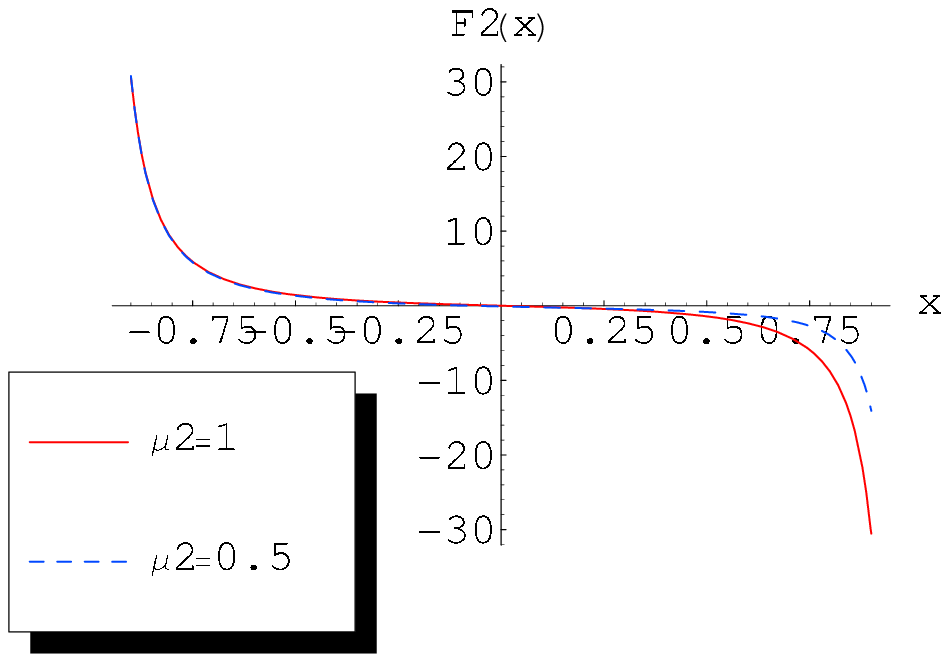,width=12cm}
\end{center}
\caption{Force as a function of the position of the test particle 
$x\in ]y_{1},y_{2}[$ for two delta
impurities, with $y_{2}=-y_{1}=1$ and $\mu_1=1$.\label{fig:F2xmu2}}
\end{figure}

\begin{figure}[htb]
\begin{center}
\epsfig{file=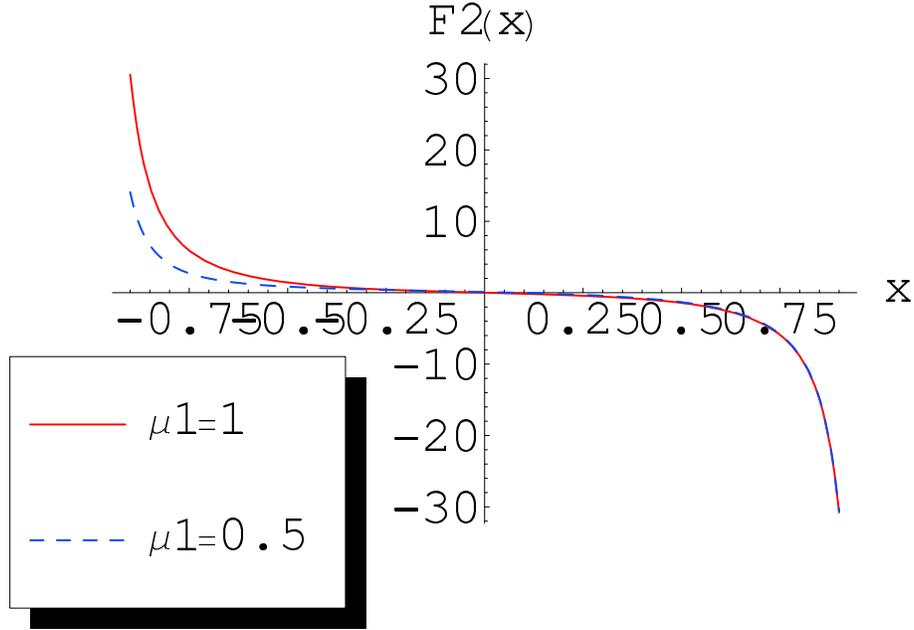,width=12cm}
\end{center}
\caption{Force as a function of the position of the test particle 
$x\in ]y_{1},y_{2}[$ for two delta
impurities, with $y_{2}=-y_{1}=1$ and $\mu_2=1$.\label{fig:F2xmu1}}
\end{figure}

Using eq. (\ref{for-cas}), one can compute the force 
on a test particle when $x\in D_{2}$, that is to say $y_{1}<x<y_{2}$. We
get
\begin{eqnarray}
\pi\,\cF_{2}(x) &=&  \int_{0}^{+\infty}dk\,\frac{k^4}{k^4+G(k)}\,\Big\{
k\Big [-\mu_1\,\cos(2k(x-y_{1}))+\mu_2\,\cos(2k(x-y_{2}))\Big] \nonu
&&\qquad\qquad\qquad\qquad\quad
+\mu_1^2\,\sin(2k(x-y_{1}))+\mu_2^2\,\sin(2k(x-y_{2}))\Big\}\, ,\nonu
G(k) &=& k^2\Big [ (\mu_1+\mu_2)^2-4\mu_1\mu_2\,\sin^2(k\,y_{12})\Big ] 
+2k\,\mu_1\mu_2(\mu_1+\mu_2)\,\sin(2k\,y_{12})\nonu
&& +4\mu_1^2\mu_2^2\,\sin^2(k\,y_{12})\, .
\nonumber
\end{eqnarray}
One can check that the integrand has no real poles, so that 
the divergent parts are obtain in the limit $k\to\infty$. They 
can be computed analytically and we obtain:
\begin{eqnarray}
\pi\,\cF_{2}(x) &=& \frac{\mu_1}{4(x-y_{1})^2} - \frac{\mu_2}{4(x-y_{2})^2}
+\frac{\mu_1^2}{2(x-y_{1})}
-\fc_{2}(x,y_{1},y_{2})-\fs_{2}(x,y_{1},y_{2})\, ,\qquad
\end{eqnarray}
where we have introduced
\begin{eqnarray}
\fc_{2}(x,y_{1},y_{2}) &=&
\int_{0}^{+\infty}dk\,\frac{k\,G(k)}{k^4+G(k)}\,\Big\{
-\mu_1\,\cos[2k(x-y_{1})]+\mu_2\,\cos[2k(x-y_{2})]\Big\}\, ,\nonu
\fs_{2}(x,y_{1},y_{2}) &=& \int_{0}^{+\infty}dk\,\frac{G(k)}{k^4+G(k)}\Big\{
\mu_1^2\,\sin[2k(x-y_{1})]+\mu_2^2\,\sin[2k(x-y_{2})]\Big\}\, .
\nonumber
\end{eqnarray}
The integral $\fc_{2}(x,y_{1},y_{2})$ has logarithmic singularities at $x=y_1,\ y_2$,  
$x=2y_1-y_2$ and $x=2y_2-y_1$, which can be extracted by means of 
the Meijer's function
$$
M(x)=G^{2\,1}_{1\,3}\left(\left.\atopn{0}{0\,0\,\half}\right\vert x^2\right)
=\int_{0}^\infty dk\,\frac{k\,\cos(2kx)}{1+k^{2}} \,.
$$
We get
\begin{eqnarray}
\fc_{2}(x,y_{1},y_{2}) &=& -\mu_{1}^{3}\,M(x-y_{1}) +\mu_{2}^{3}\,M(x-y_{2})
-\mu_{1}^{2}\,\mu_{2}\,M(x+y_{2}-2y_{1}) \nonu
&& +\mu_{1}\,\mu_{2}^{2}\,M(x+y_{1}-2y_{2})
+\wt\fc_{2}(x,y_{1},y_{2})
\end{eqnarray}
The integrals $\wt\fc_{2}(x,y_{1},y_{2})$ and 
$\fs_{2}(x,y_{1},y_{2})$ are absolutely convergent and 
can be computed numerically.
We represent the force for different values of $\mu_{1}$ and $\mu_{2}$
in figures \ref{fig:F2xmu2} and \ref{fig:F2xmu1}.

We find a point between the defects where the force changes sign. 
In this point the test particle is in stable equilibrium. 

\subsection{Force on a test particle on the left of the defects\label{sec:CasLft}}

\begin{figure}[htb]
\begin{center}
 \epsfig{file=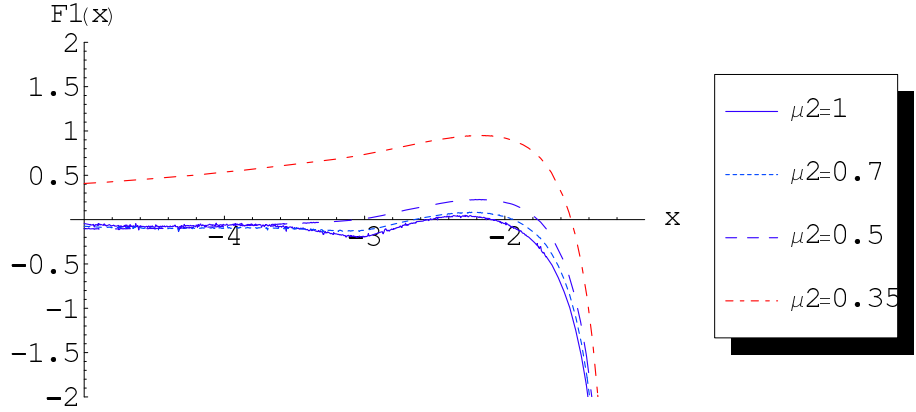,width=12cm}
\end{center}
\caption{Force as a function of the position of the test particle 
$x<y_{1}$ for two delta
defects, with $y_{2}=-y_{1}=1$ and $\mu_1=1$.\label{fig:F1xmu2}}
\end{figure}

\begin{figure}[htb]
\begin{center}
\epsfig{file=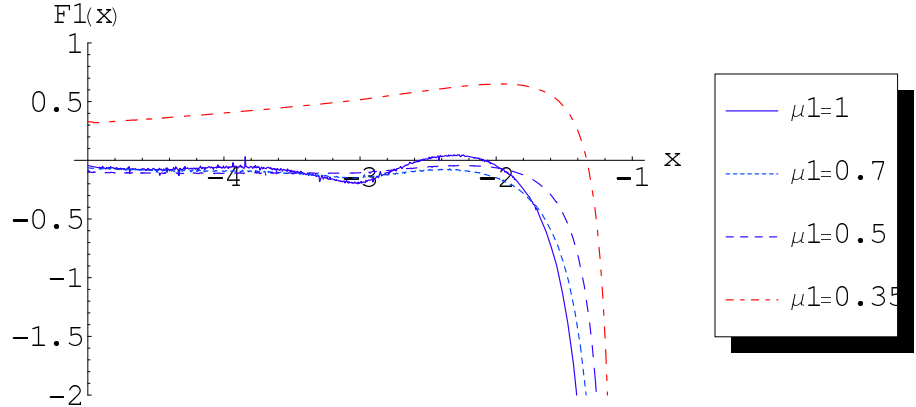,width=12cm}
\end{center}
\caption{Force as a function of the position of the test particle 
$x<y_{1}$ for two delta
defects, with $y_{2}=-y_{1}=1$ and $\mu_2=1$.\label{fig:F1xmu1}}
\end{figure}

We perform the same kind of calculation when $x\in D_{1}$, that is to say
$x<y_{1}$. Extracting the poles, we
get
\begin{eqnarray*}
\pi\,\cF_{1}(x) &=& \frac{-\mu_1}{4(x-y_{1})^2} - \frac{\mu_2}{4(x-y_{2})^2}
+\frac{\mu_1^2}{2(x-y_{1})}+\frac{\mu_2^2+2\mu_1\mu_2}{2(x-y_{2})}
\nonumber\\[1.2ex]
&&-\fc_{1}(x,y_{1},y_{2})-\fs_{1}(x,y_{1},y_{2})+
\int_{0}^{+\infty}dk\,\frac{k\,\fa_{1}(k)+\fa_{0}(k)}{k^4+G(k)}\, ,
\end{eqnarray*}
where we have introduced:
\begin{eqnarray}
\fc_{1}(x,y_{1},y_{2}) &=&
\int_{0}^{+\infty}dk\,\frac{k\,G(k)}{k^4+G(k)}
\,\Big\{
\mu_1\,\cos[2k(x-y_{1})]+\mu_2\,\cos[2k(x-y_{2})]\Big\}\, ,\nonu
\fs_{1}(x,y_{1},y_{2}) &=& \int_{0}^{+\infty}dk\,\frac{G(k)}{k^4+G(k)}\Big\{
\mu_1^2\,\sin[2k(x-y_{1})]+(\mu_2^2+2\mu_1\mu_2)\,\sin[2k(x-y_{2})]\Big\}\, ,
\nonumber
\end{eqnarray}
\begin{eqnarray}
\fa_{1}(k) &=&\!\!2\mu_1\mu_2\Big\{\mu_2\cos[2k(x-y_{1})]-\mu_2\cos[2k(x-y_{2})]
+\mu_1\sin[2k(x-y_{1})]\,\sin(2k\,y_{12})\Big\}\, ,\nonu
\fa_{0}(k) &=& 2\mu_1^2\mu_2^2\,\sin[2k(x-y_{1})]\,
\big[1-\cos(2k\,y_{12})\big]\, .\nonumber
\end{eqnarray}
Again, one can extract from $\fc_{1}(x,y_{1},y_{2})$ the Meijer's contribution 
\begin{eqnarray}
\fc_{1}(x,y_{1},y_{2}) &=& -\mu_{1}^{3}\,M(x-y_{1}) 
+\mu_{2}\big((\mu_{1}+\mu_{2})^{2}+2\,\mu_{1}^{2}\big)\,M(x-y_{2}) \nonu
&& -\mu_{1}\,\mu_{2}^{2}\,M(x+y_{1}-2y_{2})
+\wt\fc_{1}(x,y_{1},y_{2})\, . 
\end{eqnarray}
The remaining integrals are absolutely convergent and can be computed 
numerically.
We represent the force in figures \ref{fig:F1xmu2} and \ref{fig:F1xmu1}, which show that 
it is repulsive close to the defect at $y_1$, but for $x<y_1$ there are also domains 
of attraction.  

\subsection{Force at the defect}

\begin{figure}[htb]
\begin{center}
\epsfig{file=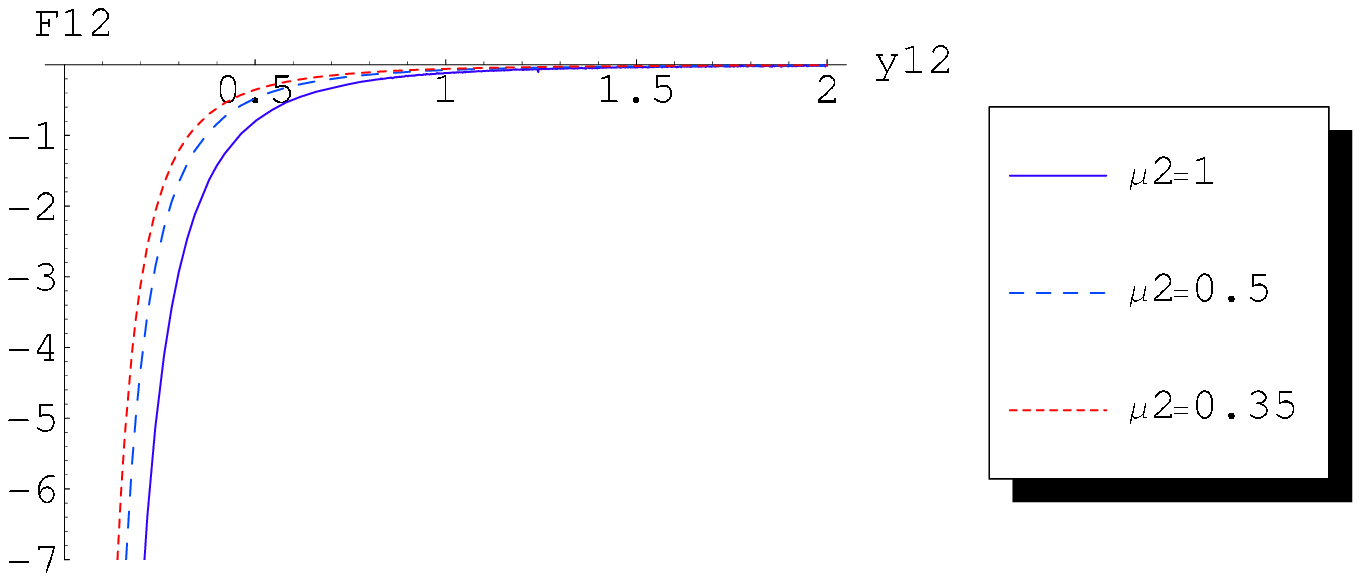,width=12cm}
\end{center}
\caption{Force at the defect for two delta
defects, as a function of the distance between them for
$\mu_1=1$.\label{fig:F12xmu2}}
\end{figure}

\begin{figure}[htb]
\begin{center}
\epsfig{file=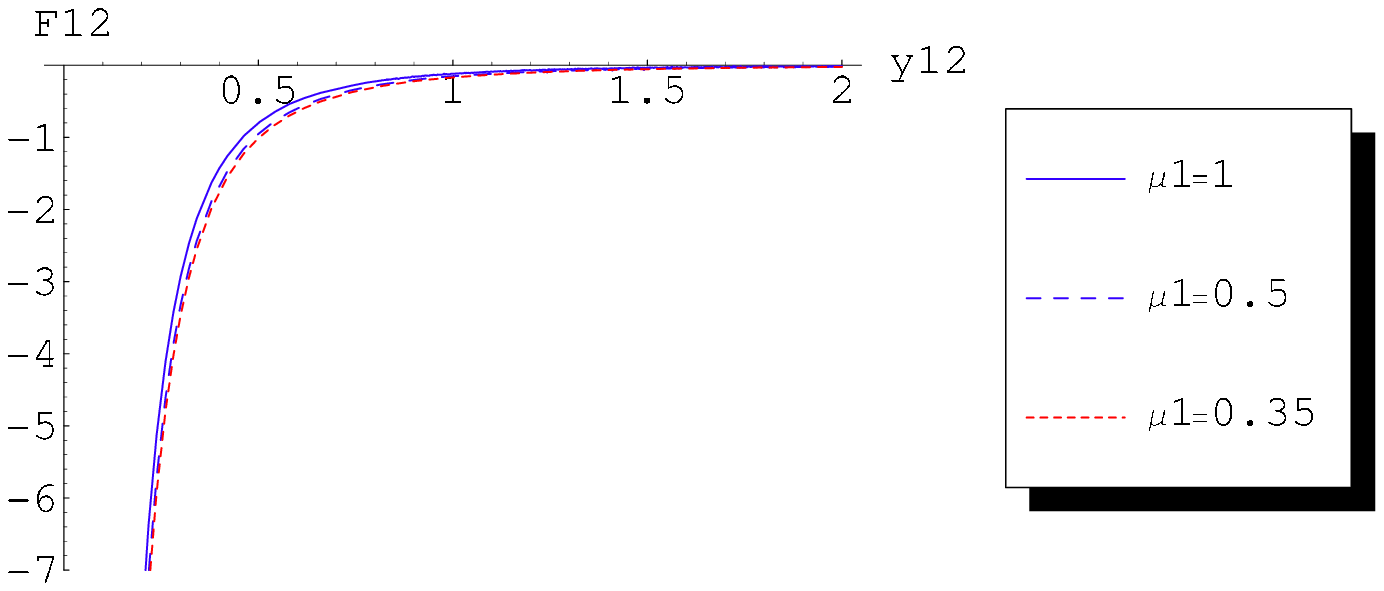,width=12cm}
\end{center}
\caption{Force at the defect for two delta
defects, as a function of the distance between them for
$\mu_2=1$.\label{fig:F12xmu1}}
\end{figure}

As already mentioned, the force is a distribution with singularities at the positions 
$x=y_1,\ y_2$ of the defects. To compute the force acting for instance at the defect in $y_{1}$, 
one needs to sum the two contributions computed in
sections \ref{sec:CasMid} and \ref{sec:CasLft} and take the limit $x\to y_{1}$, 
getting rid of the divergences. This is achieved by taking the symmetric limit:
$$
\cF_{12}=\lim_{\delta\to
0^+}\left(\cF_{1}(y_{1}-\delta)+\cF_{2}(y_{1}+\delta)\right)\,.
$$
One obtains the force as a function of the distance between the two 
defects:
\begin{eqnarray}
\pi\,\cF_{12} &=& -\frac{\mu_2}{2y_{12}^2}+\frac{\mu_2(\mu_2+\mu_1)}{y_{12}}
-2\mu_2(\mu_1+\mu_2)\,\int_{0}^{+\infty}dk\,
\frac{G(k)}{k^4+G(k)}\,\sin(2k\,y_{12})\nonu
&&-2\mu_1\,\mu_2^2\,\int_{0}^{+\infty}\,
\frac{k\,dk}{k^4+G(k)}\Big[(\mu_1+\mu_2)\,k\sin(4y_{12})
+4\mu_1\mu_2 \cos(2ky_{12})\sin^2(ky_{12})\Big]\nonu
&&-2\mu_2\,\int_{0}^{+\infty}dk\,
\frac{k^{3}}{k^4+G(k)}\Big[\mu_1\mu_2\cos(4ky_{12})
+(\mu_1^2+\mu_2^2+\mu_1\mu_2)\cos(2ky_{12})\Big]\, .
\nonumber
\end{eqnarray}
Using the Meijer's function, one finds for the last integral 
$$
-2\mu_{1}\mu_{2}^{2}\,M(2y_{12})
-2\mu_2\,(\mu_1^2+\mu_2^2+\mu_1\mu_2)\,M(y_{12})+\fc_{12}(y_{12},\mu_{1},\mu_{2})\, ,
$$ 
where $\fc_{12}(y_{12},\mu_{1},\mu_{2})$ is absolutely convergent. 
Computing the integrals numerically, one gets the figures 
\ref{fig:F12xmu2} and \ref{fig:F12xmu1}.

\subsection{Charge density}

The charged density for a complex field and two delta-defects is 
obtained from (\ref{cd2}). The integrals can be computed numerically, the 
mass $m$ and the temperature $\frac{1}{\beta}$ playing the role of 
regulators at $k=0$ and $k=\infty$ respectively. We get
$$
\wh\rho_{j}(x,\beta)=\rho_{j}(x,\beta)-\int_{-\infty}^{\infty} \frac{dk}{\pi}
\frac{\e^{-\beta \omega(k)}} { 1-\e^{-\beta \omega(k)}}
=\int_{-\infty}^{\infty} \frac{dk}{\pi}
\frac{\e^{-\beta \omega(k)}} { 1-\e^{-\beta \omega(k)}}\,
\frac{\fr_{j}(x)}{k^4+G(k)}\ ,\quad j=1,2, 
$$
with
\begin{eqnarray*}
\fr_{2}(k) &=& 2\,\Big\{ \mu_1\sin [2 k (x-y_1)] -k \mu_2 \sin [2 k  
(x-y_2)]\Big\}\,k^{3}\nonu
&& +2\,\Big\{\mu_1^2\cos [2 k (x-y_1)]+\mu_2^2 \cos [2 k (x-y_2)]\Big\}\,k^2\, ,
\nonu
\fr_{1}(x) &=& 2\,\Big\{\mu_1 \sin [2 k (x-y_1)]+\mu_2 \sin [2 k (x-y_2)]\Big\} 
k^3\nonu
&&-2\,\Big\{\mu_1^2\cos [2 k (x-y_1)] +\mu_2 (2 \mu_1+\mu_2) \cos [2 k
   (x-y_2)]\Big\} k^2\nonu
&&-4\,\mu_1\mu_2\sin (2 k y_{12}) \,\Big\{ \mu_1\cos [2 k (x- y_1)] + \mu_2 
\cos [ k (2x-y_1-y_2)]\Big\}  k\nonu
&&-4\,\mu_1^2\mu_2^2\,\Big[1- \cos (2 k y_{12}) \Big]\cos [2 k (x-y_1)]\, . 
\end{eqnarray*}
and a similar expression for $\fr_{3}(k)$. We remind that 
$\omega(k)=\sqrt{k^2+m^2}$.

\begin{figure}[htb]
\begin{center}
\epsfig{file=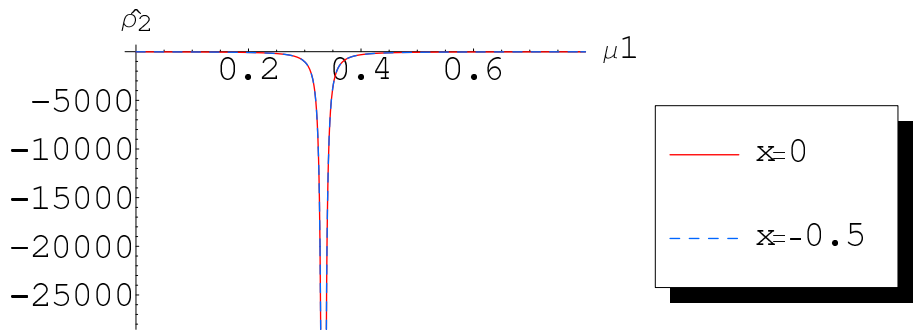,width=12cm}
\end{center}
\caption{Density between two delta
defects, as a function of  $\mu_{1}$ for
$\mu_2=1$, $y_{2}=1=-y_{1}$ and $\beta=1$.\label{fig:D2-13}}
\end{figure}

When $m=0$, there is a divergence at $k=0$ reflecting the possibility of Bose 
condensation. When $m\neq0$ (and $\beta\neq0$), the integral defining $\rho_{j}(x,\beta)$, 
$j=1$ or 3, converges. The same is true for $\rho_{2}(x,\beta)$ and 
generic values of the parameters $\mu_{1}$, $\mu_{2}$, $y_{1}$ and 
$y_{2}$. However, when the relation 
$$
\mu_{1}+\mu_{2}=2\,\mu_{1}\,\mu_{2}\,(y_{2}-y_{1})\,, 
$$
is fulfilled, $\rho_{2}(x,\beta)$ diverges (see figure \ref{fig:D2-13}). This relation corresponds 
to a change in the behavior of $G(k)$ around $k=0$:
$$
G(k) \raisebox{-1.42ex}{$\atopn{\mbox{ \Large{$\sim$} }}{\mbox{\scriptsize{$k=0$}}}$} 
\big(\mu_{1}+\mu_{2}-2\,\mu_{1}\,\mu_{2}\,y_{21}\big)^2 \,k^2+ o(k^4)\, ,
$$
and could be interpreted as a kind of resonance effect.
One can check that, at this point in the parameter space, the 
integrals entering in $\rho_{j}(x,\beta)$, 
$j=1$ or 3, as well as in the force studied in the previous sections, 
still converge.

\begin{figure}[htb]
\begin{center}
\epsfig{file=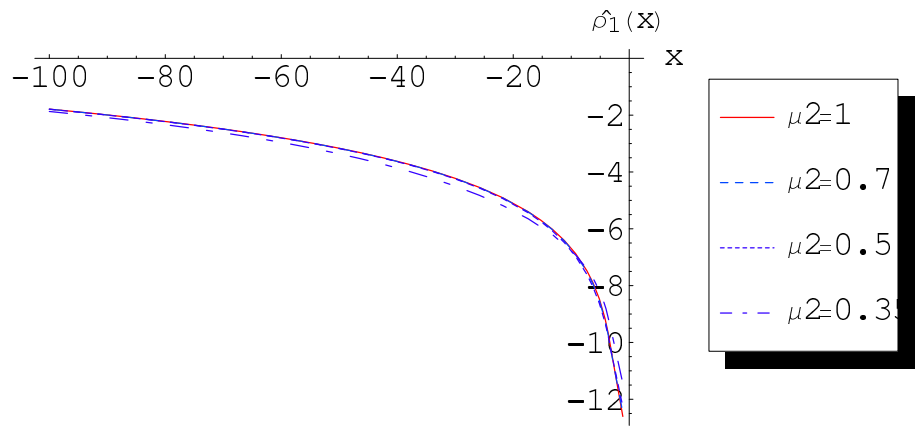,width=12cm}
\end{center}
\caption{Density on the left of two delta
defects, as a function of  $x$ for
$\mu_1=1$, $y_{2}=1=-y_{1}$ and $\beta=1$ (large distance behavior).\label{fig:D1muL}}
\end{figure}

\begin{figure}[htb]
\begin{center}
\epsfig{file=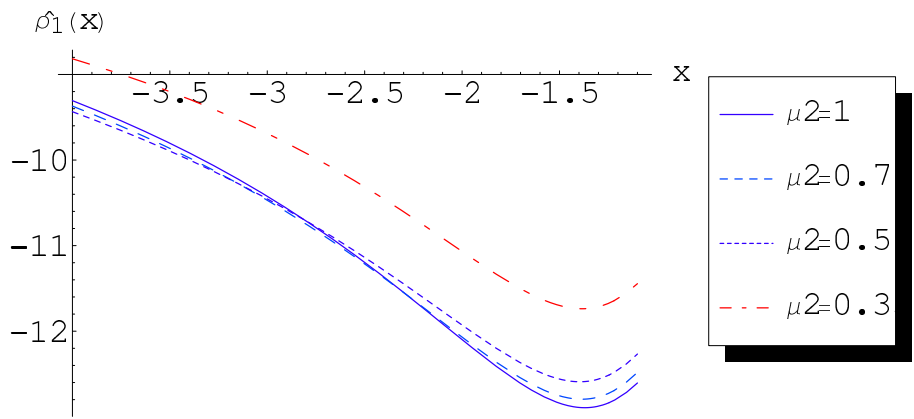,width=12cm}
\end{center}
\caption{Density on the left of two delta
defects, as a function of  $x$ for
$\mu_1=1$, $y_{2}=1=-y_{1}$ and $\beta=1$ 
(zoom of figure \ref{fig:D1muL}, close to the defect in y1=-1).\label{fig:D1muZ}}
\end{figure}

\begin{figure}[htb]
\begin{center}
\epsfig{file=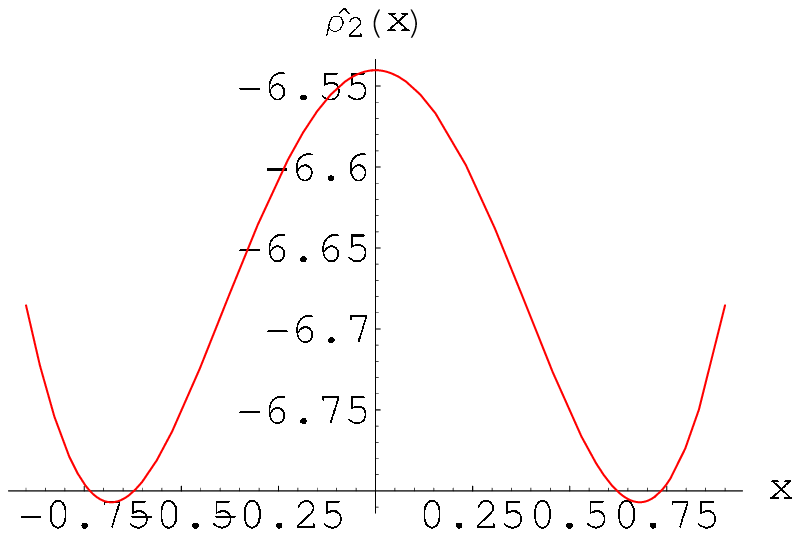,width=7cm}
\qquad\qquad
\epsfig{file=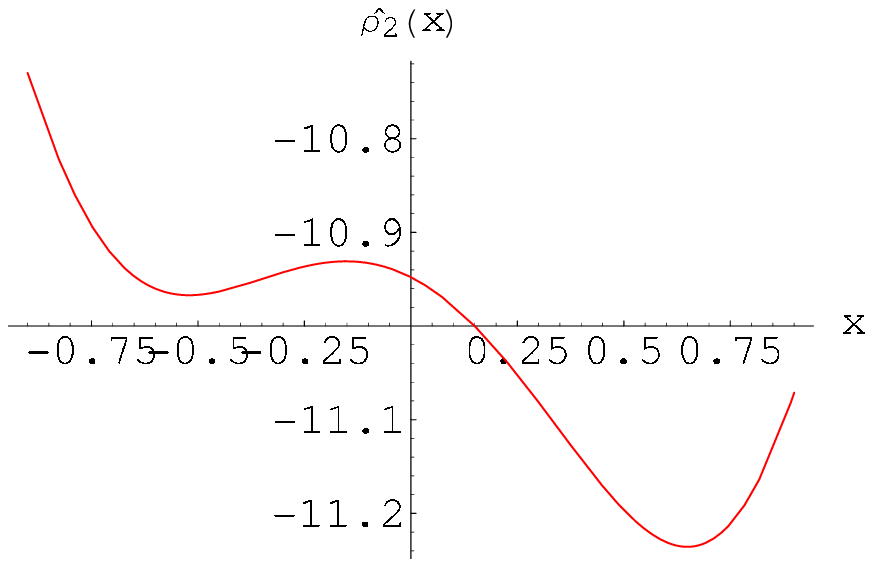,width=7cm}
\\
\hfill $\mu_{1}=1$ \hfill\hfill $\mu_{1}=0.8$\hfill\null\\
\null\ \\
\epsfig{file=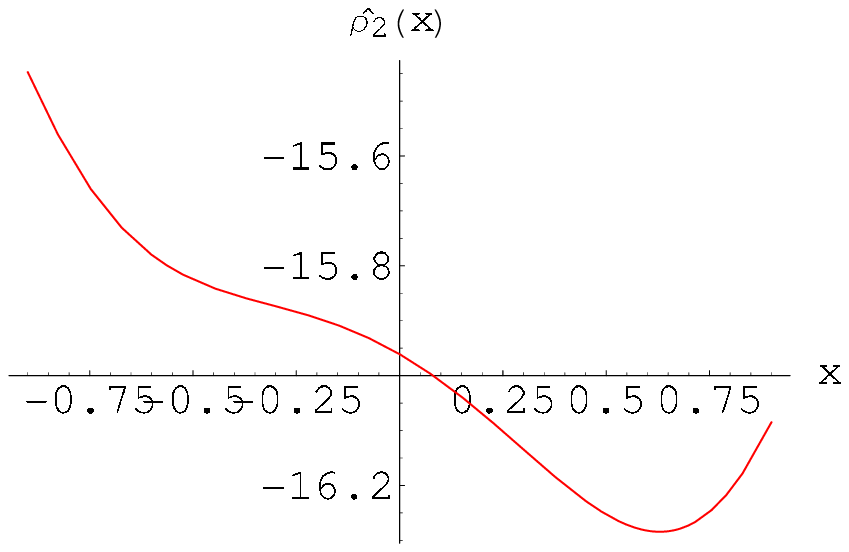,width=7cm}
\qquad\qquad
\epsfig{file=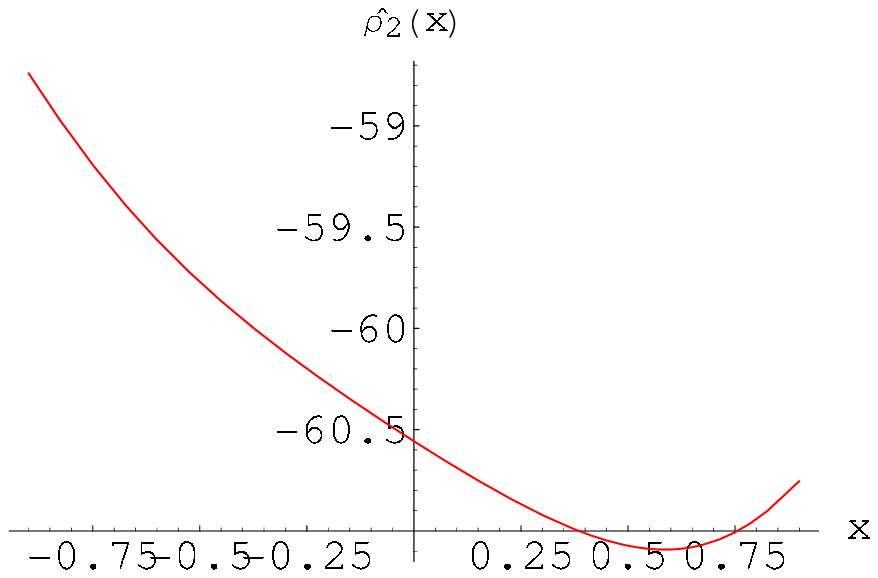,width=7cm}
\\
\hfill $\mu_{1}=0.7$ \hfill\hfill $\mu_{1}=0.5$\hfill\null\\
\null\ \\
\epsfig{file=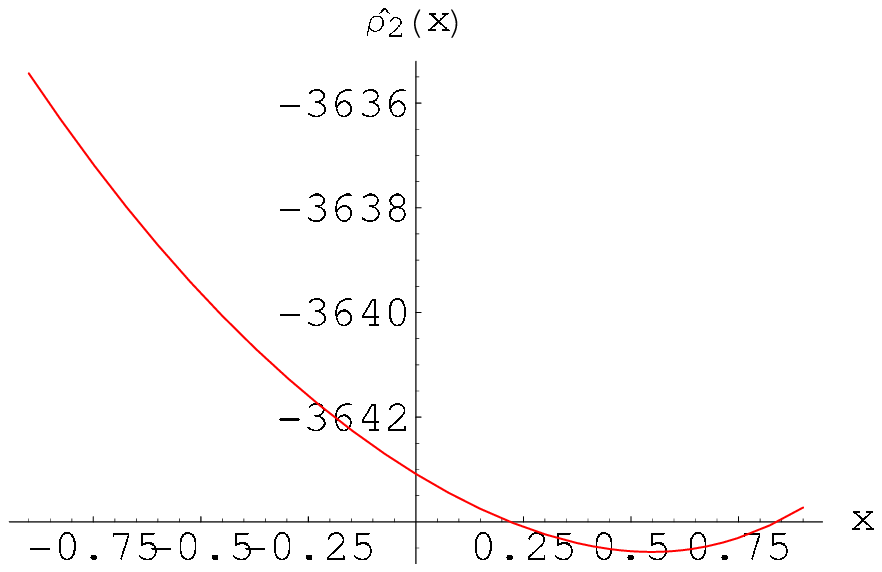,width=7cm}
\qquad\qquad
\epsfig{file=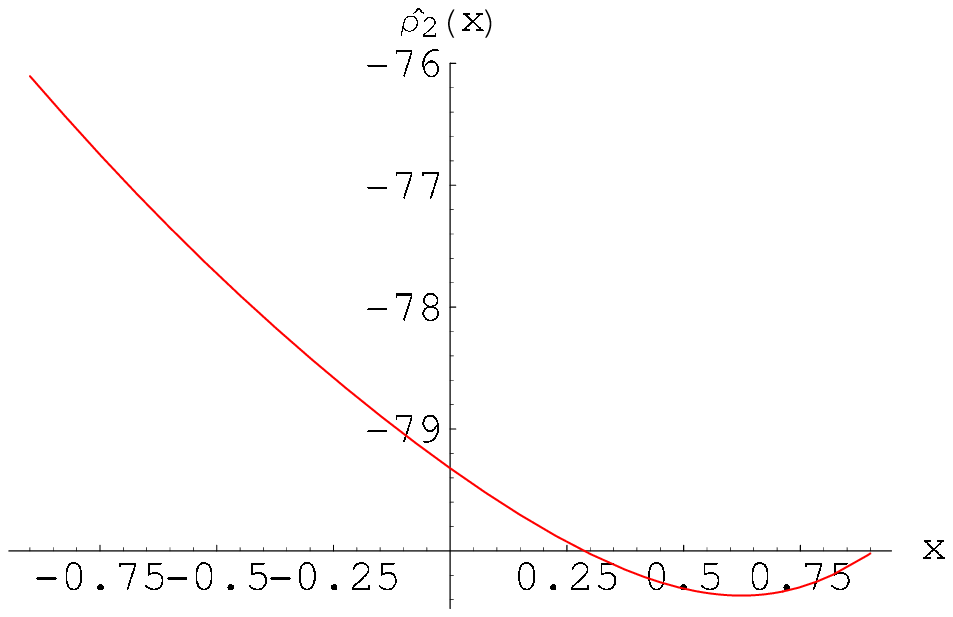,width=7cm}
\\
\hfill $\mu_{1}=0.35$ \hfill\hfill $\mu_{1}=0.2$\hfill\null\\
\end{center}
\caption{Density between two delta
defects, as a function of  $x$ for 
$\mu_2=1$ and $\beta=1$.\label{fig:D2xmu021}}
\end{figure}

The plots of $\wh\rho_{j}(x,\beta)$ are presented in figures 
\ref{fig:D1muL}-\ref{fig:D2xmu021}.
 
\clearpage

\section{Conclusion\label{sec:conclu}} 

In the present paper we generalized the RT algebra framework 
\cite{Mintchev:2002zd}-\cite{Caudrelier:2004hj} for 
dealing with one defect to any number of defects. 
More precisely, we considered a scalar field, which interacts with multiple 
defects on the line and freely propagates away of them.  
We studied the most general point-like defects, parametrizing the interaction by local  
reflection and transmission coefficients. As expected, the algebraic formulation 
considerably simplifies the
boundary value problem at hand and is very efficient for deriving the correlation
functions both at zero and finite temperature. Applied to the Casimir effect, the
algebraic technique leads to a nice and compact expression (in terms of the reflection
coefficients) for the force. As an explicit illustration, we derived the Casimir force for 
two delta-type defects. It turns out that both the intensity and the direction of the force 
depend on the position. We established also the charge density distribution of a complex 
scalar field interacting with the defects.  

Our framework can be generalized in several directions. 
One can apply it to the case where the field $\varphi$ 
has $N>1$ internal degrees of freedom. Then, each of the operators 
$a_{i}(k)$, $i=1,\ldots,n$, becomes a vector of length $N$, while 
the `coefficients' $C^{[\alpha]}_{ij}$, $i,j=1,2$ are now $N\times N$ 
matrices. The above results remain valid if these matrices commute among them.  

Another very attractive application concerns 
quantum graphs \cite{K1, K2}, 
which are networks of wires connected at nodes. 
Each node is characterized by a scattering matrix \cite{Kostrykin:1998gz} 
and can be viewed as a defect \cite{Bellazzini:2006jb}. 
The RT algebra approach has been
already applied \cite{Bellazzini:2006kh} 
to the study of quantum fields on graphs with
one vertex (star graphs). Using the results of this paper, 
one can formulate quantum field theory on a generic quantum graph and 
address in this framework the interesting problems of conductance and 
vacuum energy.  

\section*{Acknowledgments}
E.R. warmly thanks E. Pilon and JP. Guillet for useful advice on the 
analytical and numerical (resp.) estimates of the integrals 
computed in section \ref{sec:delta}.



\begin{thebibliography}{99} 

\bibitem{KF}
C.~L.~Kane and M.~P.~A.~Fisher, 
Phys.\ Rev.\ B {\bf 46} (1992) 15233. 

\bibitem{NFLL}
C.~Nayak, M.~P.~A.~Fisher, A.~W.~W.~Ludwig and H.~H.~Lin, 
Phys.\ Rev.\ B {\bf 59} (1999) 15694.

\bibitem{Sch} 
X.~Barnabe-Theriault, A.~Sedeki, V.~Meden, K.~Sch\"onhammer
Phys.\ Rev.\ B {\bf 71} (2005) 205327.

\bibitem{Oshikawa:2005fh} 
M.~Oshikawa, C.~Chamon and I.~Affleck, 
J.\ Stat.\ Mech.\  {\bf 0602} (2006) P008 
[arXiv:cond-mat/0509675].

\bibitem{Fradkin:2006mb}
E.~Fradkin and J.~E.~Moore, 
Phys.\ Rev.\ Lett.\  {\bf 97} (2006) 050404 
[arXiv:cond-mat/0605683].

\bibitem{Saleur:1998hq}
H.~Saleur,
``Lectures on Non-perturbative field theory and quantum impurity  problems",
arXiv:cond-mat/9812110.

\bibitem{Saleur:2000gp}
H.~Saleur,
``Lectures on Non-perturbative field theory and quantum impurity  problems II'',
arXiv:cond-mat/0007309.

\bibitem{Cherednik:jt}
I.~Cherednik,
Int.\ J.\ Mod.\ Phys.\ A {\bf 7} (1992) 109.

\bibitem{Delfino:1994nr}
G.~Delfino, G.~Mussardo and P.~Simonetti,
Nucl.\ Phys.\ B {\bf 432} (1994) 518
[arXiv:hep-th/9409076].

\bibitem{Konik:1997gx}
R.~Konik and A.~LeClair,
Nucl.\ Phys.\ B {\bf 538} (1999) 587
[arXiv:hep-th/9703085].

\bibitem{Castro-Alvaredo:2002dj}
O.~Castro-Alvaredo and A.~Fring,
Nucl.\ Phys.\ B {\bf 649} (2003) 449
[arXiv:hep-th/0205076].

\bibitem{Bowcock:2003dr} 
P.~Bowcock, E.~Corrigan and C.~Zambon, 
Int.\ J.\ Mod.\ Phys.\  A {\bf 19S2} (2004) 82
[arXiv:hep-th/0305022].

\bibitem{Bowcock:2004my}
P.~Bowcock, E.~Corrigan and C.~Zambon, 
JHEP {\bf 0401} (2004) 056
[arXiv:hep-th/0401020].

\bibitem{Mintchev:2002zd}
M.~Mintchev, E.~Ragoucy and P.~Sorba,
Phys.\ Lett.\ B {\bf 547} (2002) 313.
[arXiv:hep-th/0209052].

\bibitem{Mintchev:2003ue}
M.~Mintchev, E.~Ragoucy and P.~Sorba, 
J.\ Phys.\ A {\bf 36} (2003) 10407
[arXiv:hep-th/0303187].

\bibitem{Caudrelier:2004hj}
V.~Caudrelier, M.~Mintchev, E.~Ragoucy and P.~Sorba,
J.\ Phys.\ A  {\bf 38} (2005) 3431
[arXiv:hep-th/0412159].

\bibitem{Mintchev:2004jy}
M.~Mintchev and P.~Sorba,
JSTAT {\bf 0407} (2004) P001
[arXiv:hep-th/0405264].

\bibitem{Mintchev:2005rz} 
M.~Mintchev and P.~Sorba,
Annales Henri Poincare {\bf 7} (2006) 1375
[arXiv:hep-th/0511162].

\bibitem{A} S. Albeverio, L. Dabrowski and P. Kurasov, 
Lett. Math. Phys. {\bf 45} (1998) 33. 

\bibitem{SCH} A. G. M. Schmidt, B. K. Cheng and M. G. E. da Luz, 
Phys. Rev. A {\bf 66} (2002) 062712. 

\bibitem{K1}
P.~Kuchment, 
Waves Random Media {\bf 12} (2002) R1. 

\bibitem{K2}
P.~Kuchment, 
Waves Random Media {\bf 14} (2004) S107. 

\bibitem{Kostrykin:1998gz}
V.~Kostrykin and R.~Schrader, 
J.\ Phys.\ A {\bf 32} (1999) 595.

\bibitem{Bellazzini:2006jb}
B.~Bellazzini and M.~Mintchev, 
J.\ Phys.\ A  {\bf 39} (2006) 11101 
[arXiv:hep-th/0605036].
 
\bibitem{Bellazzini:2006kh}
B.~Bellazzini, M.~Mintchev and P.~Sorba,
J.\ Phys.\ A  {\bf 40} (2007) 2485
[arXiv:hep-th/0611090].


\end{thebibliography}
\end{document}